\documentstyle[aps,preprint,epsf,multicol]{revtex}

\def\be{\begin{equation}}
\def\ee{\end{equation}}
\def\beq{\begin{equationarray}}
\def\eeq{\end{equationarray}}
\def\prb{Phys. Rev. B }
\def\prl{Phys. Rev. Lett. }
\begin{document}
\draft
\title{Transmission and reflection in a perfectly amplifying and absorbing medium}
\author{P. K. Datta\cite{mail}}
\address{Mehta Research Institute of Mathematics and Mathematical Physics, Chhatnag
Road, Jhusi, Allahabad-211 019, India}
\maketitle
\begin{abstract} 
We study transmission and reflection properties of a perfectly
amplifying as well as absorbing medium analytically by using the tight
binding equation. Different  expressions for transmittance and reflectance
are obtained for even and odd values of the sample length which is the origin
of their oscillatory behavior. In a weak amplifying medium, a
cross-over length scale exists below which transmittance and reflectance
increase exponentially and above which transmittance decays
exponentially while the reflectance
gets saturated. Depending on amplification transmittance and
reflectance show singular behavior at the
cross-over length. In a weak
absorbing medium we do not find any cross-over length
scale. The transmission coefficient behaves similar to that in an
amplifying medium in the asymptotic limit. In a strong
amplifying/absorbing medium, the transmission coefficient decays
exponentially in the large length limit. In both weak as well as strong
absorbing media the logarithm of the reflection coefficient shows the
same behavior as obtained in an amplifying medium but with opposite
sign. 

\end{abstract}

In recent years the study of wave propagation in an active random medium,
in the presence of absorption or amplification has attracted much
interest [1-23]. Due to the absence of a conservation law for photons,
light may be absorbed
or amplified in the medium while phase coherence is preserved. The
interplay of absorption or amplification and localization has been
studied extensively by using the Helmholtz equation with an imaginary
dielectric constant of an appropriate sign or by using the Schr\"odinger
equation with imaginary potentials. Experimentally a random amplifying
medium can be achieved in a turbid laser dye or a powdered laser
crystal \cite{law}. The optical propagation or magnetic excitons in solids which
terminate upon reaching trapping sites can be thought of as an absorbing
medium. Several results have been obtained in this area. The system plays a
dual role as an
amplifier or absorber and a reflector \cite{gup}. When the strength of the
imaginary potential is increased beyond a certain limit both absorber
and amplifying scatterer act as a reflector. Absorption suppresses
the transmission just as amplification
does in the large length limit \cite{pas}. A cross-over length scale exists below which the
amplification enhances the transmission and above which it reduces the
transmission \cite{pas,zha,jos,jos1}. In contrast, the reflectance
saturates in the large length
limit for both types of media \cite{jos,jos1}. The distribution of
reflectance \cite{bee,bro},
transmittance \cite{gup,zha} and the phase of transmittance \cite{fre2} and
reflectance \cite{gup,jos} for both
the active media has been studied by several authors analytically
\cite{pra,fre,fre3,hen} as
well as numerically \cite{gup,jos,jos1,kim}.

Most of the studies have been made on active random media. For proper
understanding of the effect of randomness in active media it is
important to study the perfect absorbing as well as amplifying
system. But much less attention has been paid in this direction 
\cite{zek,jayan,sen}. So, we
study here the transmission and reflection properties of both the
perfect systems by using the tight binding equation,
\be
(E - \epsilon_n) c_n = V (c_{n-1} + c_{n+1}).\label{tbe}
\ee
Here, $E$ is the particle energy, $V$ is nearest-neighbor hopping
amplitude, $\epsilon_n$ is the $n$-th site energy. and $c_n$ is the
amplitude of the $n$-th site. Without any loss of generality we assume
$V = 1$. For a perfectly active medium $\epsilon_n = i \eta$ makes the
Hamiltonian non-Hermitian and consequently the particle conservation
fails. Here $\eta$ is a fixed real number which is positive (negative)
for an absorbing (amplifying) medium. In a previous study \cite{sen}
it has been
shown numerically with some analytical calculations that for a perfectly
absorbing medium transmittance decays
exponentially with the sample length but reflectance saturates after
initial oscillations. On the other hand, for an amplifying medium
transmittance initially increases up to a certain length through large
oscillations after which it decays and the 
reflectance also shows an initial increasing behavior after which it saturates
to a value larger than unity. In Ref.\cite{zek} a perfectly amplifying medium
has been studied in the
Kronig Penny model. The authors obtain exact expressions for transmission
coefficient and the cross-over length. They have also studied the
disorder effect on the system. However, we feel that the work on
perfect amplifying/absorbing medium is not complete. Some interesting
results which may help to understand the properties of active media in the
presence of randomness seem to have been overlooked in previous
studies \cite{zek,jayan,sen}. So,
here we perform the analytical study of perfect amplifying as well as
absorbing media using Eq. \ref{tbe}.  By using the transfer
matrix method we obtain exact expressions for the 
transmission and reflection coefficients for both the media, described
by Eq. \ref{tbe}. The oscillatory behavior of both transmittance and
reflectance is understood from those
expressions. Depending on values of $\eta$ for the amplifying medium we
find singular behavior in the transmission and reflection coefficients
at the cross-over length. For both amplifying and absorbing media, the
transmittance and reflectance are studied in different ranges of $\eta$
and sample length. 

We now discuss the transfer matrix method \cite{pkd} to calculate
the transmission and reflection coefficients. The active medium
consisting of $N$ sites ($n = 1$ to $N$) is placed between two
semi-infinite perfect leads with all site-energies are taken to be zero and
$V = 1$. From Eq. \ref{tbe} one can easily obtain the
site-amplitude for any length of the chain from the initial ones
through sequential product of transfer matrices in the following
way,
\be
\left( \begin{array}{l}
c_{N+1}\\
c_{N}\end{array}\right)
=\prod_{i=1}^N P_i
\left( \begin{array}{l}
c_{1}\\
c_{0} \end{array}\right)\label{trf}
\end{equation}
where the transfer matrix
$$P_i = 
\left( \begin{array}{ll} 
E - \epsilon_i & -1\\
1                  &  0\end{array}\right).
$$
Note also that \(P_i\) is a unimodular matrix. If a plane wave
$e^{i k n}$ with energy $E = 2 \cos k$ is sent through the perfect lead
from one side then the
solutions on the two sides of the sample are related by
$$M = \Omega S^{-1} \prod_{i=1}^{N} P_{i} S$$
where
$$
\Omega = 
\left( \begin{array}{ll} 
e^{-i k N} & 0\\
0                  &  e^{i k N}\end{array}\right),~~~~
S = \left ( \begin{array}{cc} e^{ik} & e^{-ik} \\ 1 & 1 \end{array} \right ).
$$
The transmission amplitude can be written as
\be
t = \frac{1}{M_{22}} = \frac{-2 i \sin k \;e^{-i k N}}{(B - C) + (A
e^{-i k} -D e^{i k})}\label{tra}
\ee
and the reflection amplitude 
\be
r = - \frac{M_{21}}{M_{22}} = \frac{e^{i k}(D - A) + (C e^{2 i k} -
B)}{(B - C) +(Ae^{-ik} - D e^{i k})}\label{rfa}
\ee
where $A, B, C$ and $D$ are four matrix elements of $\prod_{i = 1}^{N}
P_i$. All the four matrix elements have different expressions for even
and odd values of $N$ which are given in the table 1. In our study we
set the energy of the incident
particle at $E = 0$ i.e. at the midband energy. Replacing $A, B, C$
and $D$ in Eq. \ref{tra}
and \ref{rfa} we get the exact expression for transmission and reflection
coefficients for both even and odd values of $N$ as
\be
T_{\rm even} = |t|^2 = \frac{\cosh^2 \xi}{(\sinh N \xi + \cosh \xi \;
\cosh N \xi)^2}\label{trne}
\ee
\be
T_{\rm odd} =  \frac{\cosh^2 \xi}{(\cosh N \xi + \cosh \xi \;
\sinh N \xi)^2}\label{trno}
\ee
\be
R_{\rm even} = |r|^2 = \frac{\sinh^2 \xi \sinh^2 N \xi}{(\sinh N \xi +
\cosh \xi \; \cosh N \xi)^2}\label{refe}
\ee
and
\be
R_{\rm odd} = \frac{\sinh^2 \xi \cosh^2 N \xi}{(\cosh N \xi + \cosh
\xi \; \sinh N \xi)^2}\label{refo}
\ee
where $\sinh \xi = \eta/2$.
Subscripts even and odd indicate for even and odd values of $N$
respectively. In Figs. 1-3 we have plotted $\ln T$ and
$\ln R$ for weak media as a
function of $N$ by
using the above equations and from numerical calculations using
Eqs. \ref{tra} and \ref{rfa}. It is found that the
oscillations arise due to the different behavior of transmission and
reflection coefficients for even and odd values of $N$. Using the
above equations we next
discuss the behavior of transmission and reflection coefficients for
both amplifying and absorbing media at different regions of $N$ for
both small and large values of $|\eta|$.

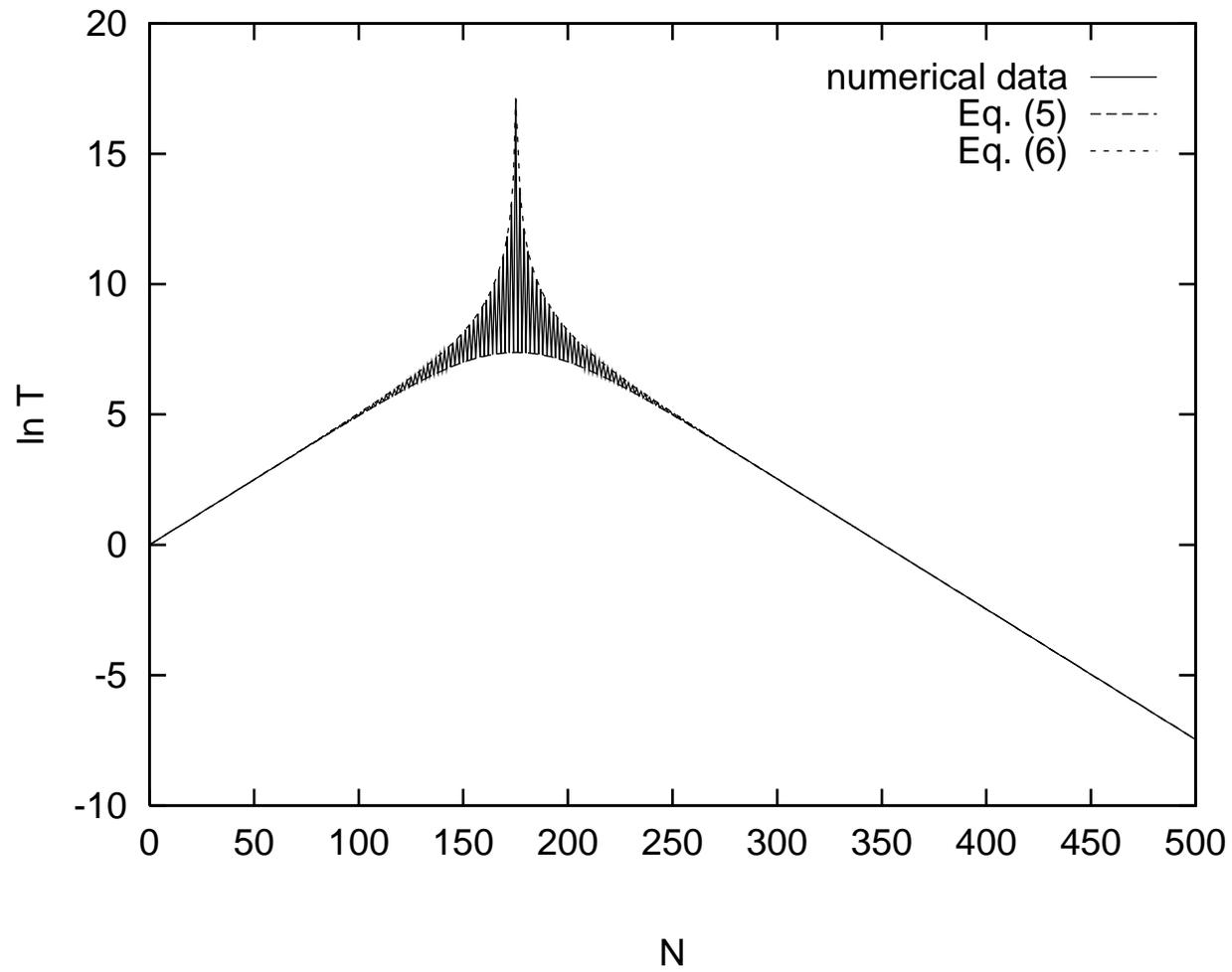
\begin{figure}
\caption{Plot of $\ln T$ as a function of sample length $N$ for an
amplifying medium with $\eta = -0.05$.}
\label{tam}
\end{figure}

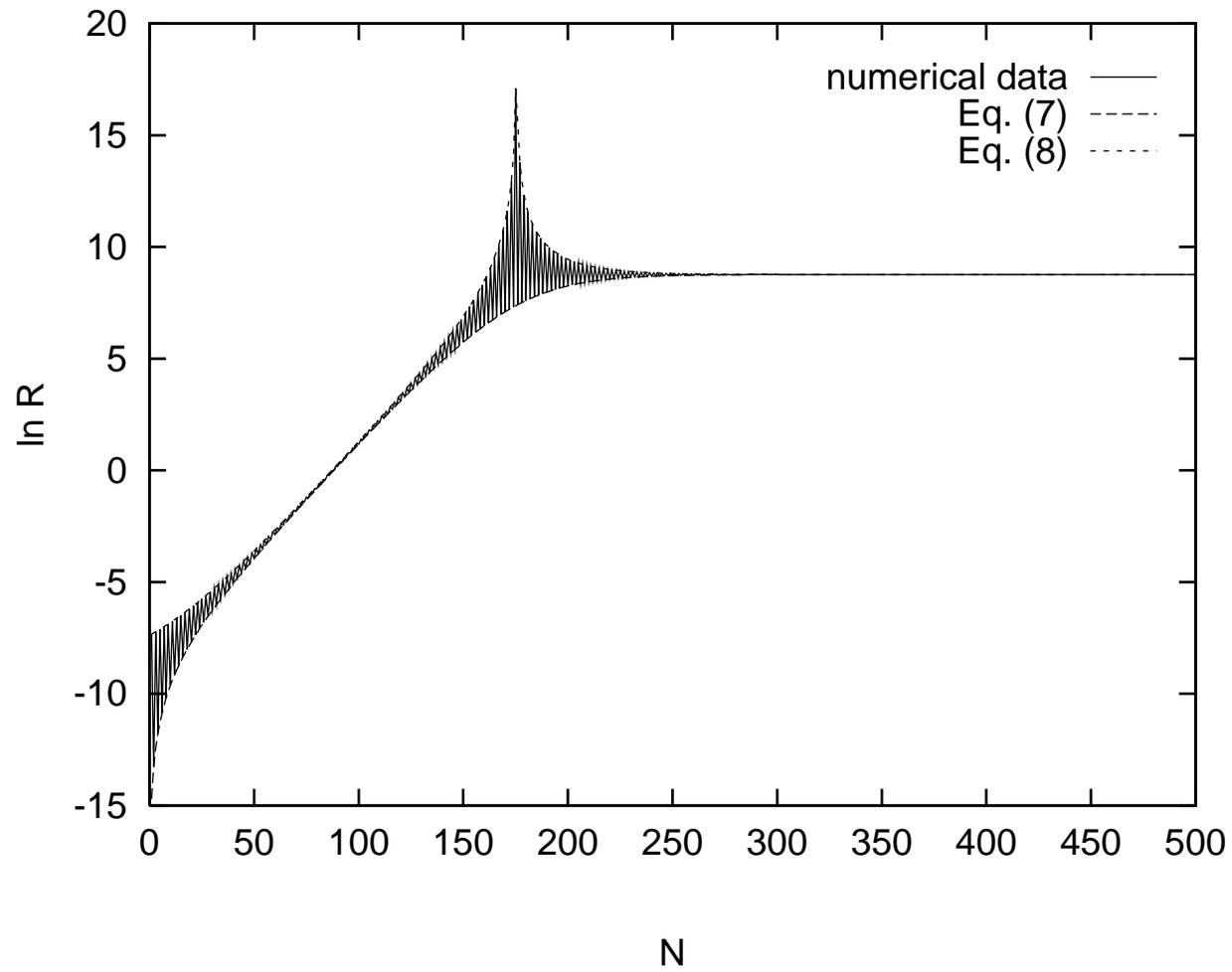
\begin{figure}
\caption{Plot of $\ln R$ as a function of $N$ for the same system as in
Fig. \ref{tam}.}
\label{ram}
\end{figure}

We first consider the case of weak amplifying medium i.e. $-1 <<\eta
<0$ and $N << N_{\rm cr} = \frac{4 \ln 2 - 2 \ln |\eta|}{|\eta|}$. The
transmission coefficient $(T)$ shows same behavior for both odd and even values
of $N$ as
\be
\ln T \approx |\eta| N.
\ee
Thus, below $N_{\rm cr}$ the transmission coefficient
increases in the weak amplifying medium exponentially (see
Fig. \ref{tam}) with the rate
$|\eta|$ \cite{fre,pas}. For fixed value of $N (<<N_{\rm cr})$ as we increase
$|\eta|$ the transmission coefficient increases. This behavior also sustains
in the presence of randomness in the medium \cite{jos1}. The reflection
coefficient $(R)$ shows different behavior for even and odd values of
$N$ as
\be
\ln R_{\rm even} \approx 2 \ln \frac{|\eta|}{4} + 2 \ln(e^{N |\eta|}-1)
\ee
and
\be   
\ln R_{\rm odd} \approx 2 \ln \frac{|\eta|}{4} + 2 \ln(e^{N |\eta|}+1).
\ee
>From the above two expressions it can be easily understood that the
difference between $\ln R_{\rm even}$ and $\ln R_{\rm odd}$ is
significant for small values of $N$. This is also observed in Fig. \ref{ram}
where the oscillations occur in $\ln R$ (obtained from numerical
calculations) for small
values of $N$. When $N$ is large but much less than $N_{\rm cr}$ the reflection
coefficient increases exponentially with $N$ with the rate of $2
|\eta|$. On the other hand for a fixed value of $N (<< N_{\rm cr})$ as
we increase $|\eta|$ both $T$ and $R$ increase. Thus in this limit
the amplification enhances the transmission as
well as reflection coefficients. 

In the asymptotic limit of $N$ (i.e. $N >> N_{\rm cr}$) and for weak
amplification the transmission and
reflection coefficients do not depend on even and odd values of
$N$. They are,
\be
\ln T \approx 4 \ln \frac{4}{|\eta|} -|\eta| N
\ee
and
\be
\ln R \approx 2 \ln \frac{4}{|\eta|}.
\ee
The transmission coefficient decreases exponentially (see
Fig. \ref{tam}) with $N$ at
the rate of $|\eta|$. Thus for $N >> N_{\rm cr}$ amplification
suppresses the transmission with the localization length $N_0 =
1/|\eta|$ \cite{pas,zha}. It should be noted that in both the regions $N
<< N_{\rm cr}$ and $N >> N_{\rm cr}$ the rate of increase and
decrease of $T$ is equal.  The reflection coefficient attains to a
saturation value (see Fig. \ref{ram}) depending on $|\eta|$. This value
increases as $|\eta|$ decreases.      

In the case of weak amplification $T$ increases for small values of $N$
and decreases in the asymptotic
limit. So, there exists a length scale, called the cross-over
length scale where the transmission
coefficient shows a maximum. Our next job is to find the cross-over length
scale. The
transmission coefficient will reach its maximum value when the first
derivative of $T$ with respect to $N$ vanishes. From Eq. \ref{trne}
 we find that $T_{\rm even}$ shows a
maximum at 
  \be
  N_{\rm max} = \frac{1}{2 \xi} \ln \left (\frac{\cosh \xi - 1}{\cosh \xi
+ 1}\right ). \label{maxe}
\ee
>From the above expression we find that the positive value of $N_{\rm
max}$ is obtained only when $\eta < 0$ i.e. the peak of $T$ occurs in an
amplifying medium \cite{pas,zha,zek,jos1}. On the other hand $T_{\rm odd}$
goes to infinity as
$N \rightarrow N_{\rm max}$. This is also true for $R_{\rm odd}$ as
the denominator of both $R$ and $T$ is same. The active medium is
described here as a discrete system containing $N$ amplifying/absorbing
sites where $N$ takes only even and odd integer values. So, with proper
choice of $\eta$
if $N_{\max}$ becomes odd integer we obtain the singular behavior of
transmission coefficient. In Fig. \ref{maxt} we have plotted the values
of $\ln T_{\rm max}$ (i.e. maximum value of $T$) obtained
from numerical calculations as a function of
$|\eta|$ which contains many sharp peaks indicating the presence of
singularity of $T$ at different values of $|\eta|$. This kind of
oscillatory behavior is also observed around the peak of the plot of
$\ln T_{\rm max}$ in the presence of randomness in
the medium (see
Fig.[1] in Ref. \cite{jos}). All the singularities in $T$ are
destroyed by the randomness in the system. The randomness in the
medium suppresses the weak amplification. Consequently, the amplitude
of oscillations of $\ln T_{\rm max}$ decreases with $|\eta|$
decreasing. However, in the perfectly weak
amplifying
medium the position of the singularities in $T$ are shown in
Fig. \ref{maxn} (obtained from numerical calculations). For comparison
we also plot Eq. \ref{maxe}. The
deviation occurs at large values of $|\eta|$ due to the discreteness of
the system. For any value of $|\eta|$ near $N_{\rm max}$, $T_{\rm odd}$
 diverges rapidly to a very large value (infinity,
depending upon
$\eta$) whereas $T_{\rm even}$ shows a finite
maximum value. This indicates that the transmission coefficient 
oscillates around $N_{\rm max}$. As the system is discrete we obtain a
limit of $\eta$ to get a maximum in $T$. We find that $\eta = -2$ when
$N_{\rm max} = 1$. Thus the maximum in
$T$ occurs only when $0 > \eta > -2$. In the
case of weak amplifying medium $(-1<< \eta <0)$ we find $N_{\rm max} =
N_{\rm cr}$.

For large values of $|\eta|$ (i.e. $|\eta| >> 2$) and in the asymptotic
limit of $N$ the transmission and reflection coefficients show the
same behavior 
for even and odd values of $N$. The transmission
coefficient decays exponentially with $N$ as,
\be
\ln T \approx 2 \ln 2 + 4/|\eta|+4/|\eta|^2 - 2 (\ln |\eta|
+1/|\eta|^2) N. \label{tams}
\ee
It should be noted that in the strong amplifying medium the rate of
decay is different from that in a weak amplifying medium. The reflection
coefficient attains to a saturation value which depends on $\eta$ as
\be
\ln R \approx 4/|\eta| - 8/3 |\eta|^3.\label{rams}
\ee
Thus as $|\eta|$ increases reflection coefficient in saturation region
decreases and approaches to unity. On the other hand $T$ decreases
exponentially with $N$. This implies
that for larger values of $|\eta|$ the medium behaves as a
reflector.

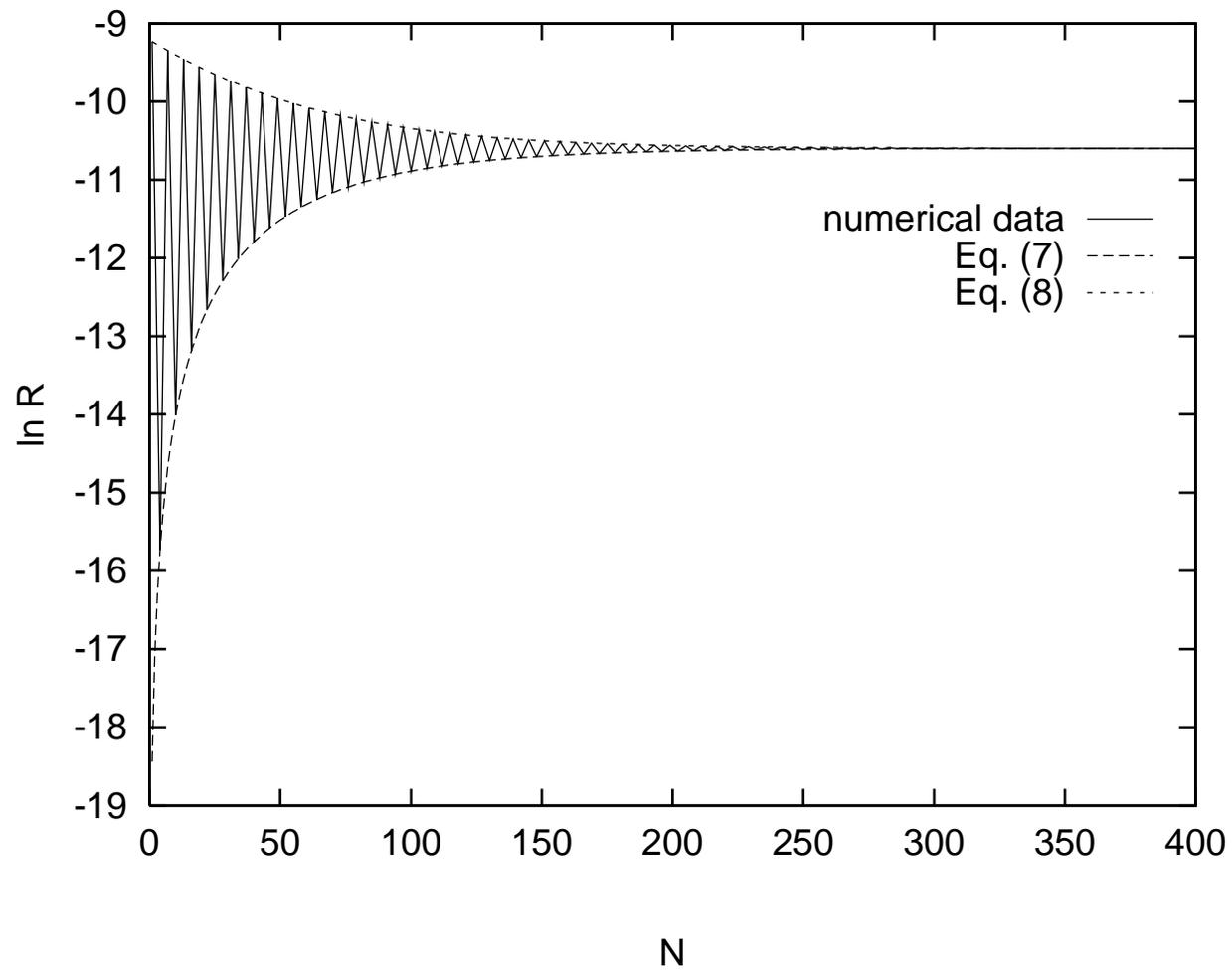
\begin{figure}
\caption{Same as Fig. \ref{ram} but $\eta = 0.02$}
\label{rab}
\end{figure}

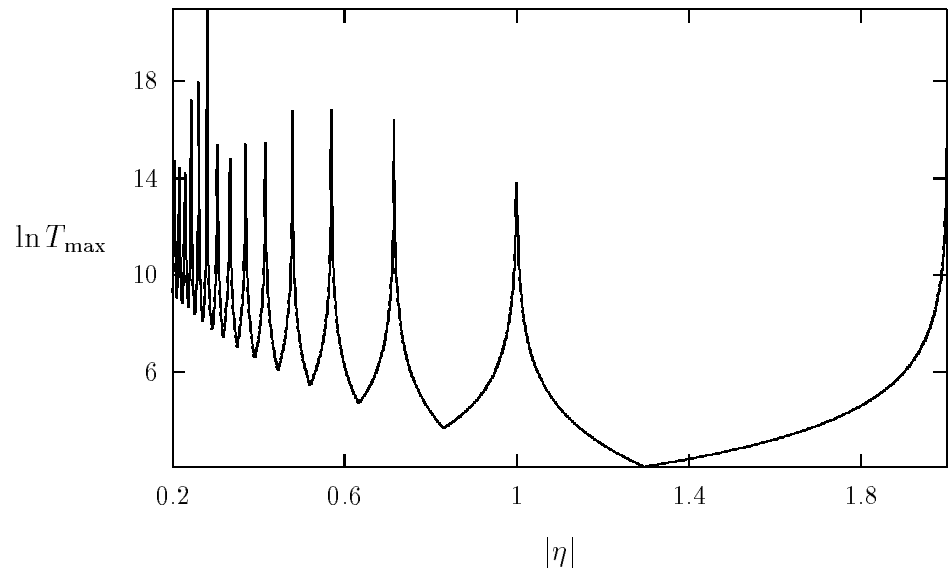
\begin{figure}
\caption{Plot of $\ln T_{\rm max}$ as a function of $|\eta|$ for an
amplifying medium.}
\label{maxt}
\end{figure}

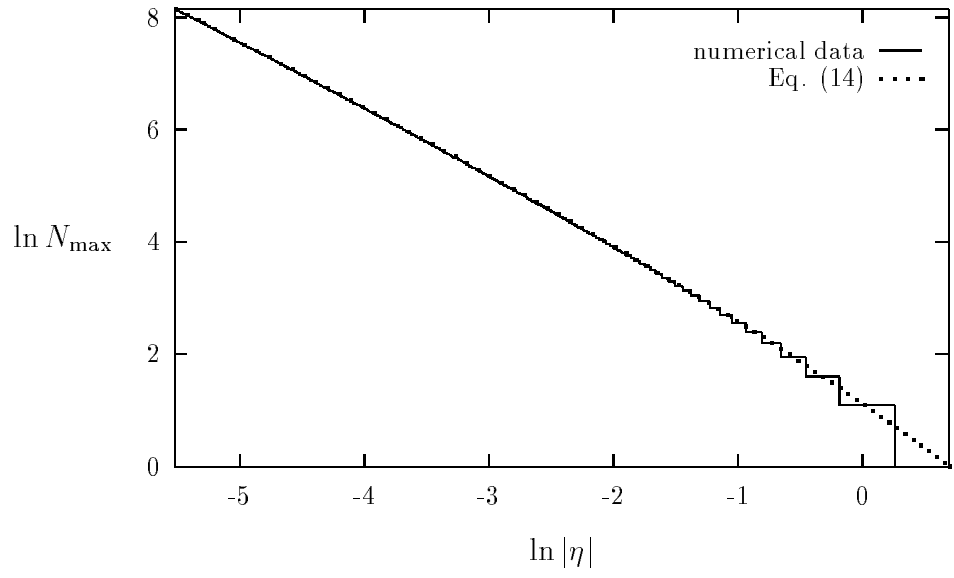
\begin{figure}
\caption{Log-log plot of $N_{\rm max}$ as a function of $|\eta|$ for 
an amplifying medium.}
\label{maxn}
\end{figure}

Now we discuss the transmission and reflection properties for an
absorbing medium
(i.e. $\eta >0$). For weak absorbing medium (i.e. $0 < \eta <<1)$ we 
find that the expression for the transmission coefficient is the same for even
and odd values of $N$ and it decays exponentially \cite{gup,kim} with $N$ as
\be
\ln T \approx -\eta N.
\ee
It should be noted that the rate of decay is the same as in a weak
amplifying medium when $N >> N_{\rm cr}$. To obtain this expression we
find that there is a restriction on the sample length of
$N >> - N_{\rm cr}$. This implies that the above expression is valid
for any values of $N (> 0)$. Hence, like in a weak amplifying medium we
do not find
any increasing behavior in $\ln T$. Though $T$ does not depend on
even and odd
values of $N$, the reflection coefficient shows different behavior as,
\be
\ln R_{\rm even} \approx 2 \ln \frac{\eta}{4} + 2 \ln(1 - e^{-\eta N})
\ee
and
\be
\ln R_{\rm odd} \approx 2 \ln \frac{\eta}{4} + 2 \ln(1 + e^{-\eta N}).
\ee
The difference between the two expressions is the last term which is
significant for small values of $N$ as long as $e^{-\eta N}$ has
significant contributions. For very small values of $N$ (such that
$\eta N << 1$) the last term of $\ln R_{\rm even}$ increases with $N$
 as $\sim 2 \ln (\eta N) - \eta N$ and that of $\ln R_{\rm
odd}$ decreases as  $\sim 2 \ln 2 - \eta N$. Due to the different
behavior of $\ln R$ for small even and 
odd values of $N$ the oscillatory behavior in $R$ is obtained (see
Fig. \ref{rab}). In
asymptotic limit of $N$ we find that the reflection
coefficient is independent of $N$ and it increases with $\eta$
as $\sim \eta^2/16$. It should be noted that in the asymptotic limit of $N$ the
expression of $\ln R$ for both weak amplifying and absorbing media is the
same but with opposite sign.

For large values of $\eta$ (i.e. $\eta >> 2$) and in the asymptotic limit
of $N$ the transmission coefficient is the same for both even and odd
values of $N$ and it decays as 
\be
\ln T \approx 2 \ln 2 - 4/\eta+4/\eta^2 - 2 (\ln \eta +
1/\eta^2) N. \label{tras}
\ee
It should be noted that the rate of decay is same as that of a strong
amplifying medium (compare Eq. \ref{tams} and \ref{tras}). The
reflection coefficient attains to a saturation value which is
independent of $N$ as
\be
\ln R \approx -4/\eta +8/3 \eta^3.\label{reas}
\ee
The expression is again same as that of strong amplifying
medium but with opposite sign. It should be noted that in saturation
region reflection coefficient increases as $\eta$ increases and in the
asymptotic limit of $\eta$ the system behaves like a reflector (as $R
\rightarrow 1$) \cite{jayan}. 

In conclusion, we have studied the transmission and reflection
properties of perfectly amplifying and absorbing media by using the tight
binding equation. We obtained exact expressions for transmission and
reflection coefficients which are different for even and odd values of
$N$. This explains the origin of oscillating behavior in $\ln R$ and
$\ln T$ in different ranges of $N$. In case of an amplifying medium,
a cross-over length scale exists below which $T$ and $R$
increases exponentially and above which $T$ decreases
exponentially and $R$ attains to a saturation value. At the cross-over
length depending upon values of $\eta$  singular
behavior in $T$ and $R$ is obtained. We also found a limit of $\eta$
to get the
maximum in $\ln T$. For $|\eta| >> 2$, $T$ decreases
exponentially with the rate different from that in a weak amplifying
medium and $R$ attains to a saturation value. The expressions of $R$ in
the saturation regime are different for weak and strong amplifying
media. In case of absorbing media $T$ decays with $N$ like in an
amplifying medium in the asymptotic limit of $N$. The logarithm of the
reflection coefficient shows the same behavior as in an amplifying medium
but with opposite sign. In asymptotic limit of $|\eta|$ and $N$ both
absorbing and amplifying medium behaves like reflectors.


\newpage

\begin{table}
\caption{ The matrix elements of $\prod_{i=1}^{N} P_i$ for even
and odd values of $N$. Here $\eta = 2 \sinh \xi$.}
\vspace {0.3in}
\begin{tabular}{|c|l|l|}
\hline
$N$ & Even & Odd\\
\hline
Matrix elements &  & \\
\hline
$A$ & 
$ (-1)^{N/2} \; \frac{\cosh (N+1) \xi}{\cosh \xi} $ &
$i (-1)^{(N+1)/2} \; \frac{\sinh (N+1) \xi}{\cosh \xi}$ \\
{ } &  { } &  { }   \\
\hline
$B = -C$ &
 $i (-1)^{(N/2+1)} \; \frac{\sinh N \xi}{\cosh \xi}$ & 
$(-1)^{(N+1)/2} \; \frac{\cosh N\xi}{\cosh \xi} $ \\
{ } &  { } & { } \\
\hline
$D$ & 
$ (-1)^{N/2} \; \frac{\cosh (N-1) \xi}{\cosh \xi}$ &
 $ i (-1)^{(N+1)/2} \; \frac{\sinh (N-1) \xi}{\cosh \xi}$  \\
{ } &  { } & { } \\
\hline
\end{tabular}
\end{table}

\begin{references}
\bibitem[a]{mail} Permanent address: K. N. College, Berhampore,
Murshidabad, West Bengal-742 101, India
\bibitem{law} N. M. Lawandy, R. M. Balachandran, A. SL. Gomes and
E. Sauvain, Nature {\bf 368}, 436 (1994);
D. S. Wiersma, M. P. van Albada and Ad Lagendijk, \prl {\bf 75}, 1739 (1995).

\bibitem{wea}R. L. Weaver, \prb {\bf 47}, 1077 (1993).

\bibitem{pra} P. Pradhan and N. Kumar, \prb {\bf 50}, 9644 (1994).

\bibitem{fre} V. Freilikher, M. Pustilnik and I. Yurkevich,
Phys. Rev. Lett. {\bf 73}, 810 (1994)

\bibitem{gup} A. K. Gupta and A. M. Jayannavar, \prb {\bf 52}, 4156 (1995).

\bibitem{pas} J. C. J. Paasschens, T. Sh. Misipashaev and C. W. J. Beenakker,
\prb {\bf 54}, 11887 (1996).

\bibitem{zha} Z. Q. Zhang, \prb {\bf 52}, 7960 (1995).

\bibitem{bee} C. W. J. Beenakker, J. C. J. Paasschens and P. W. Brouwer, \prl
{\bf 76}, 1368 (1996).

\bibitem{zek} N. Zekri, H. Bahlouli and A. K. Sen,
J. Phys.:Condens. Matter {\bf 10}, 2405 (1998).

\bibitem{mis} T. Sh. Misirpashaev, J. C. J. Paasschens and C. W. J. Beenakker,
Physica A {\bf 236}, 189 (1997).

\bibitem{yos} M. Yosefin, Europhys. Lett. {\bf 25}, 675 (1994).

\bibitem{joh} S. John, \prl {\bf 53}, 2169 (1984).

\bibitem{gen} A. Z. Genack, \prl {\bf 58}, 2043 (1986).

\bibitem{gen1} A. Z. Genack and N. Garcia, \prl {\bf 66}, 2064 (1991).

\bibitem{fre2} V. Freilikher and M. Pustilnik, \prb {\bf 55}, R653
(1997).

\bibitem{fre3} V. Freilikher, M. Pustilnik, and I. Yurkevich,
\prb {\bf 56}, 5974 (1997).

\bibitem{jos} S. K. Joshi and A. M. Jayannavar, \prb {\bf 56},
12038 (1998).

\bibitem{jos1} S. K. Joshi and A. M. Jayannavar, cond-mat/9712250.

\bibitem{kim} K. Kim, cond-mat/980334.

\bibitem{den} W. Deng, D. S. Wiersma, Z. Q. Zhang, \prb {\bf 56}, 178
(1997).

\bibitem{bro} P. W. Brouwer, \prb {\bf 57}, 10526 (1998).

\bibitem{hen} J. Heinrichs, \prb {\bf 56}, 8674 (1997).

\bibitem{bur} A. A. Burkov and A. Yu. Zyuzin, \prb {\bf 55}, 5736
(1997).

\bibitem{jayan} A. M. Jayannavar, \prb {\bf 49}, 14718 (1994). 

\bibitem{sen} A. K. Sen, Mod. Phys. Lett. B {\bf 10}, 125 (1996).

\bibitem{pkd} P. K. Datta, D. Giri and K. Kundu, \prb {\bf 47}, 10727 (1993).

\end{references}
\end{document}